\documentclass[aip,jap,twocolumn,reprint,showpacs,floatfix]{revtex4-1}

\providecommand{\e}[1]{\ensuremath{\times 10^{#1}}}
\usepackage{graphicx}% Include figure files
\usepackage{dcolumn}% Align table columns on decimal point1
\usepackage{bm}% bold math
\usepackage{subfigure}  % use for side-by-side figures

\begin{document}

%\title{Ultracold ion source temperature measurement with a time-dependent electric field}
\title{Measurement of the temperature of an ultracold ion source using time-dependent electric fields}

\author{N. Debernardi, M.P. Reijnders, W.J. Engelen, T.T.J. Clevis, P.H.A. Mutsaers, O.J. Luiten and E.J.D. Vredenbregt}
\email{E.J.D.Vredenbregt@tue.nl}
\affiliation{Department of Applied Physics, Eindhoven University of Technology, P.O Box 513, 5600 MB Eindhoven, the Netherlands}

\date{\today{}}

\begin{abstract}
%The source temperature, $(1 \pm 2)$ mK, and the corresponding transversal reduced emittance, $7.9 \e{-9}$ m rad $\sqrt{eV}$, have been experimentally measured with a new and direct method. The effect of space charge has been studied at low energies (few tens of eV) and at ultra low average charge per bunch (0.02 ions per bunch). A new way to focus ion beams with time-dependent electric fields is presented in this paper.
We report on a measurement of the characteristic temperature of an ultracold rubidium ion source, in which a cloud of laser-cooled atoms is converted to ions by photo-ionization. Extracted ion pulses are focused on a detector with a pulsed-field technique. The resulting experimental spot sizes are compared to particle-tracking simulations, from which an effective source temperature $T = (3 \pm 2)$ mK and the corresponding transversal reduced emittance  $\epsilon_r = 1.4 \e{-8}$ m rad $\sqrt{\rm{eV}}$ are determined. Space charge effects that may affect the measurement are also discussed. %We find that this result is likely limited by space charge forces even though the average number of ions per bunch is 0.022.
\end{abstract}


\maketitle

\section{Introduction}
Focused Ion Beams (FIBs) are widely utilized in the semiconductor industry and in nanoscience \cite{Orloff_Book_03, Orloff_RoSI_93}. They are used successfully for high precision milling or deposition and on the other hand, similarly to scanning electron microscopy (SEM), for high resolution microscopy \cite{Ward_JVSTB_06}. To keep up with reduction in sizes in the semiconductor industry it is necessary to reduce the smallest focusable spot size \cite{itrs, Moore_E_65}. The properties of the source have a key role in what can be achieved. A typical FIB using the current industry-standard Liquid Metal Ion source (LMIS) can reach a high brightness ($10^6$ A m$^{-2}$ sr$^{-1}$ eV$^{-1}$) and can deliver 10 pA current in 10 nm spot size \cite{Orloff_RoSI_93} (note that higher currents can be reached if the beam is less focused). Brightness is the current per unit area and solid angle, normalized by the beam energy and it is a key property for a source. A new kind of ion source, the ultracold ion source (UCIS) has been proposed as an alternative for the LMIS \cite{Hanssen_PRA_06, Geer_JoAP_07, Claessens_PoP_07}. The UCIS is based on creating very cold ion beams by near-threshold photo-ionization of a laser-cooled and trapped atomic gas, with a source temperature $T$ less than 1 mK. The UCIS has the potential of producing ion beams with a brightness of $10^5$ A m$^{-2}$ sr$^{-1}$ eV$^{-1}$ and a current up to 100 pA, according to simulations \cite{Geer_JoAP_07}. Its major advantage is an energy spread two orders of magnitude lower than the LMIS (down to 10 meV) as demonstrated by Reijnders \textit{et al.}\cite{Reijnders_PRL_09}. Lower energy spread may lead to a smaller achievable spot size by reducing the contribution of chromatic aberration\cite{Geer_JoAP_07}.\\ %When the chromatic aberration is the dominant term, this could result a 0.8 nm spot size, with an usable current of 1 pA \cite{Geer_JoAP_07, Kruit_NL_08}.\\
This work aims to measure the source temperature of a rubidium UCIS, which is an important physical quantity related to the ability to focus the beam. Here we show how an effective source temperature can be extracted from measurements of the minimum spot size achieved in focusing the beam and we demonstrate a new way to focus ion beams. The ultra low temperature of the source permits collimated beams to be created at low energy (down to a few eV \cite{Reijnders_PRL_09}), which allows using time-dependent fields for accelerating and focusing. The duration and the shape of the accelerating electric field pulse can be tuned so that a variable focusing lens is created. An effective source temperature is then extracted from waist scans varying the focal strength of the time-dependent lens.

\section{Principle of the measurement}
\label{sec:prin}
\begin{figure}[h]
    \centering
        \includegraphics[width=1\linewidth]{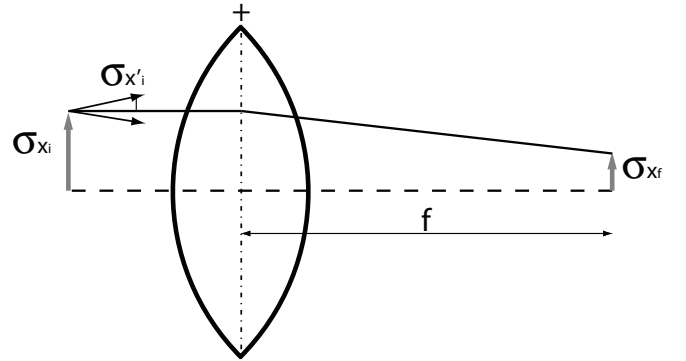}
    \caption{Principle of the source temperature measurement. Ion bunches are focused on a detector with a time-dependent lens with focal length $f$. The quantity $\sigma_{x'_i}$ is the initial \textit{rms} angular spread, $\sigma_{x_i}$ is the initial \textit{rms} radius of the source and $\sigma_{x_f}$ the \textit{rms} radius of the final spot on the detector.}
    \label{fig:opt_model}
\end{figure}

In the experiment, ion bunches, extracted from the source with longitudinal energy $U$, are focused on a detector with a variable focal length $f$. With a simple first order optical model it is possible to find an upper limit for the source temperature from the final root-mean-square (\textit{rms}) spot radius $\sigma_{x_f}$ (see Fig. \ref{fig:opt_model} for a schematic drawing), assuming that the emittance is conserved.

The size of the ion bunch at the detector position is completely determined by the initial \textit{rms} angular spread $\sigma_{x'_i}$ when the image is in focus,
\begin{equation}
    \sigma_{x_f} = f \sigma_{x'_1}.
    \label{eq:angSpread}
\end{equation}
The \textit{rms} value of the angular spread is related to the source temperature $T$ by
\begin{equation}
    \sigma_{x'_i} = \sqrt{\frac{k_b T}{2U}},
    \label{eq:spread}
\end{equation}
where $k_b$ is Boltzmann`s constant and $U$ is the energy of the bunch. 
Reversing this equation, the source temperature can be defined given a measurement of $\sigma_{x_f}$ as
\begin{equation}
    T = \frac{2}{k_b} U \left(\frac{\sigma_{x_f}}{f}\right)^2.
    \label{eqT}
\end{equation} 
Any emittance growth, e.g., due to distortions in the imaging system, will affect the extracted value of $T$, which thus becomes an effective source temperature.
Knowing $T$ and thus $\sigma_{x'_i}$, we can derive the \textit{rms} normalized emittance $\epsilon_r$ of the source when, in addition, the source radius $\sigma_{x_i}$ is known. The normalized emittance is a measure of the phase space occupied by the beam multiplied by the square root of the beam energy and is given by \cite{Luiten_IJoMPA_07}
\begin{equation}
    \epsilon_r = \sigma_{x_i} \sigma_{x_f} \sqrt{U} = \sigma_{x_i} \sqrt{\frac{k_b T}{2}}.
    \label{eq:emitt2}
\end{equation}
Indeed in Eq. (\ref{eq:emitt2}) the dependence on the energy $U$ disappears indicating that the normalized emittance is a property of the source.

\section{Experimental Setup}
\label{sec:setup}
\begin{figure}[h]
    \centering
        \includegraphics[width=1\linewidth]{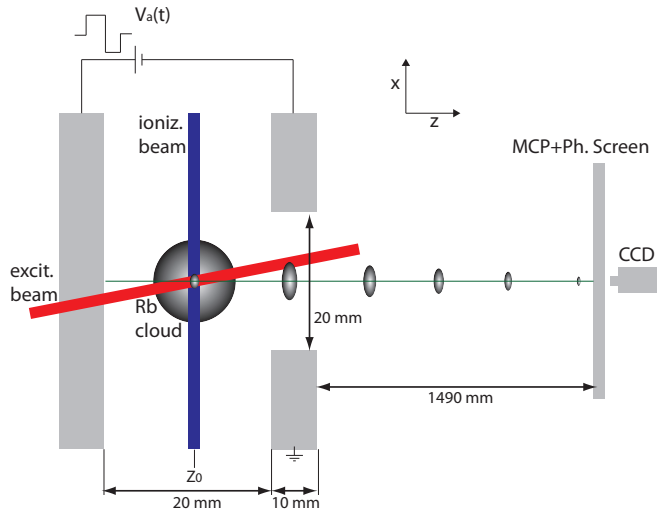}
    \caption{Schematic view of the experimental setup (not to scale) in the x-z plane. The ionization laser beam and the excitation laser beam select a portion of the Rb atomic cloud trapped inside the accelerator. The bunched ions are accelerated using a time-dependent potential $V_a (t)$ and after a flight distance of 1.51 m they reach a MCP detector with a phosphor screen. The images are captured by a CCD camera.}
    \label{fig:setup}
\end{figure}

The UCIS is based on the technique of laser cooling and trapping of neutral atoms\cite{Metcalf_Book_99, Foot_Book_05}. We start with a magneto-optical trap (MOT) of rubidium atoms. Three orthogonal pairs of counter propagating 780 nm laser beams (trapping laser beams) are used to Doppler cool a $^{85}$Rb atomic cloud and a quadrupole magnetic field is added to trap the atoms. Typically $10^8$ Rb-atoms are trapped in a volume of  16 mm$^3$ \textit{rms} at an expected source temperature of $T_0= 143$ $\mu$K \cite{Metcalf_Book_99}, which corresponds to 9 neV of average kinetic energy of the atoms. The MOT is surrounded by an accelerator \cite{Taban_PRSAB_08}.\\
The accelerator has a cylindrically symmetric structure and it is placed in a vacuum chamber where the rubidium pressure is $10^{-9}$ mbar. See Fig. \ref{fig:setup} for a schematic drawing of the experimental setup. The trapping laser beams enter through openings present in the structure. The atomic cloud is trapped in the middle of the accelerator at the intersection of the trapping laser beams, 10 mm away from the accelerator's exit, which consists of a circular hole with a diameter of 20 mm. The electric field strength at the starting point $z_0$ is 0.37 kV/cm per kV input voltage $V_a$ and $U = e V_a/2.05$, where $e$ is the elementary charge. For details of the accelerator structure see \cite{Taban_PRSAB_08}.

\begin{figure}[h]
    \centering
        \includegraphics[width=1\linewidth]{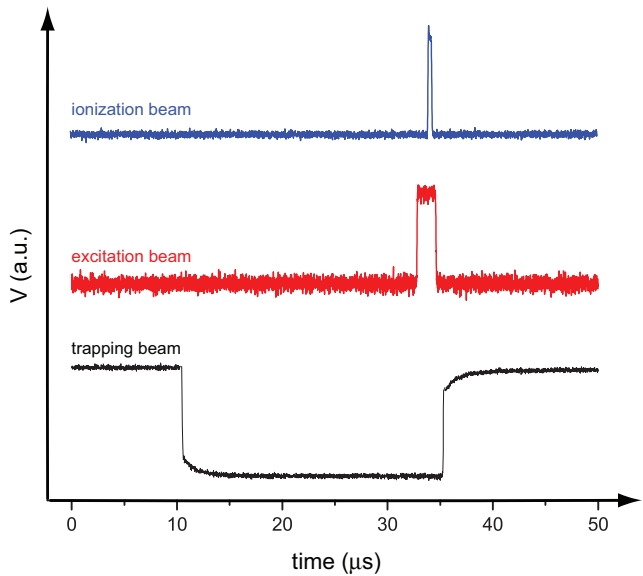}
    \caption{Typical example of the timings used in the experiment to ionize a portion of the whole atomic cloud. The trapping laser beams are turned off
for about 25 $\mu$s and during this time the excitation and ionization laser beams are turned on in coincidence. The excitation laser beam is turned on for 2 $\mu$s and the ionization laser beam for 400 ns.}
  \label{fig:timings}
\end{figure}

The ionization mechanism is a 2-step process. In order to avoid spherical aberration in the focusing process, it is necessary to work with a smaller volume than the total MOT. To achieve this, the trapping laser beams are turned off for 25 $\mu$s. During this time, a laser beam (excitation laser beam) with the same wavelength (780 nm) is focused horizontally along the z-direction to a \textit{rms} radius $\sigma_{excit} = 54$ $\mu$m. The excitation laser beam excites the Rb cloud to the 5p-level for 2 $\mu$s. Coincidentally, a 479 nm ionization laser beam is sent in vertically along the x-direction for 400 ns. The ionization laser beam also has an \textit{rms} radius $\sigma_{ioniz} = 54$ $\mu$m. Fig. \ref{fig:timings} shows an experimental example of the timing sequence used for the ionization of a portion of the Rb atomic cloud trapped in the MOT.
In this way, only the atoms at the intersection of the 2 laser beams will be ionized. The shape and position of the ionized cloud can be changed by changing the size of the lasers and their position. The position and dimension of the lasers are controlled by 2 CCD cameras, virtually positioned in the center of the MOT, which is at $z_0$ on the beam axis. The initial ionization volume is not spherical: in the x-direction the \textit{rms} radius $\sigma_x$ is determined by the intersection of two laser beams and hence
\begin{equation}
     \frac{1}{\sigma_x} = \sqrt{\frac{1}{{\sigma_{excit}}^2} + \frac{1}{{\sigma_{ioniz}}^2}},
\end{equation}
so that $\sigma_x$ is a factor $\sqrt{2}$ smaller than the size of the two laser beams.\\
The initial ionization volume size has been experimentally and numerically determined as follows. Ion bunches are created in a DC electric field and are consequentially accelerated towards the detector (which is described below). With a DC accelerating field, the exit of the accelerator forms an aperture lens with a negative focal length $f_0$ of 33 mm (``exit kick'' effect), which is independent of the acceleration voltage \cite{Taban_PRSAB_08}. This leads to a magnified image of the source on the detector with a known magnification of 46. From the final spot size, it is therefore possible to determine the source size. The initial angular spread has a negligible influence (less than $1\%$) on the final spot size for $U > 200$ eV and this measurement was performed at $U=3$ keV. The initial volume is assumed to be Gaussian distributed in all 3 directions from the fact that the profile of the lasers is Gaussian as well. Analysis of the recorded detector images give $\sigma_x =$ (38 $\pm$ 2) $\mu$m and $\sigma_y =$ (54 $\pm$ 3) $\mu$m. In the z-direction, $\sigma_z =$ 54 $\mu$m because it equals the \textit{rms} radius of the ionization laser beam $\sigma_{ioniz}$.\\
A double multichannel plate (MPC) detector with phosphor screen is located at 1.51 m from the ion's starting position $z_0$. The diameter of the MCP is 40 mm. A 16 bit, cooled CCD camera is used to image the phosphor screen through a lens placed in between with a magnification of 0.33. The spatial distribution of the ion bunches can be extracted from the CCD images. The resolution of the detector has been determined experimentally by placing two pinholes, with diameters of 25 $\mu$m and 50 $\mu$m, downstream in front of the MCP. Ion bunches with $U = 3$ keV passed through the pinhole. The \textit{rms} radius $\sigma_a$ of an aperture with diameter $D$ is equal to $D/4$. The \textit{rms} spot radius $\sigma_{det}$ measured at the detector was substantially larger than $\sigma_a$, as a result of the resolution of the detector. Multiple experiments, with both the pinholes, have been performed in order to obtain sufficient statistical accuracy. Using
\begin{equation}
     \sigma_{det} = \sqrt{\delta^2 + \sigma_a^2},
\end{equation}
the \textit{rms} resolution of the detector $\delta$ is found to be $(95 \pm 4)$ $\mu$m. In what follows, $\delta$ is quadratically subtracted from the measured spot in order to obtain $\sigma_{x_f} = \sqrt{\sigma_{det}^2 - \delta^2}$.

\section{Focusing with time-dependent electric fields}
\begin{figure}[h]
    \centering
        \includegraphics[width=1\linewidth]{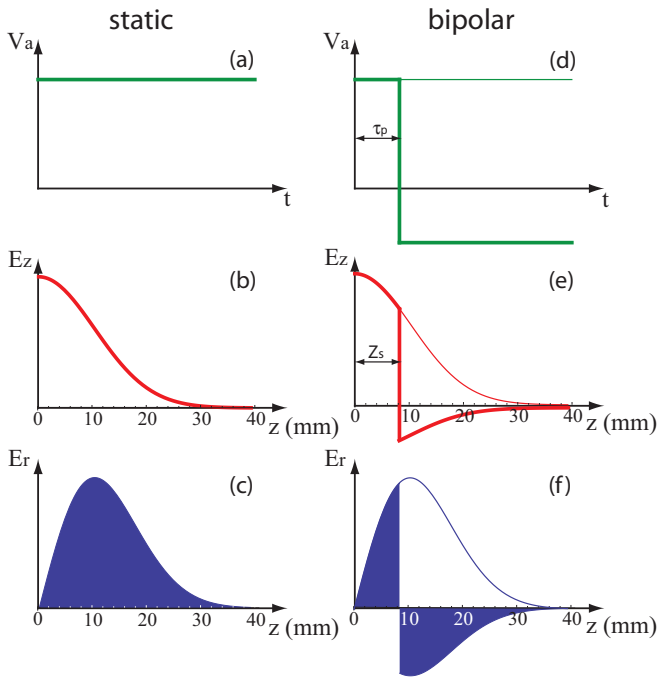}
    \caption{The potential applied to the anode $V_a$, the longitudinal electric field component $E_z$ and the radial electric field component $E_r$ (for $r \neq 0$), in case of a static (left hand side) or bipolar field (right hand side). The quantities $\tau_p$ and $z_s$ are the length respectively in time and space of the positive part of the pulse. In panels (d), (e) and (f), the thin line indicates the static case. The y-axis is in arbitrary unit in all the six
plots.}
    \label{fig:fields}
\end{figure}

\begin{figure}[h]
    \centering
        \includegraphics[width=1\linewidth]{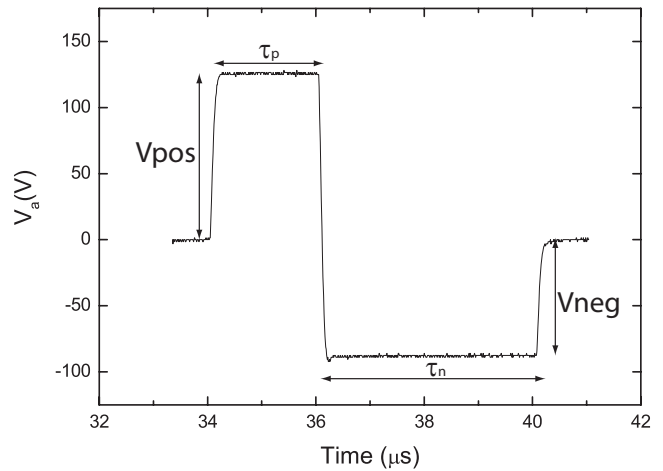}
    \caption{Example of a typical bipolar anode voltage pulse $V_a$ used in the experiment. Here, $V_{pos} = 125$ V is the positive voltage with duration $\tau_p=2$ $\mu$s and $V_{neg}= -90$ V is the negative voltage with duration $\tau_n=4$ $\mu$s. The pulse is created with a programmable waveform generator, amplified 50 times and applied to the accelerator.}
    \label{FIG:pulse}
\end{figure}

Measuring the effective temperature of the source requires focusing the ion bunches onto the detector. In fact, as shown in Section \ref{sec:prin}, at the beam waist the final spot size is completely determined by the initial angular transverse spread, which depends on the source temperature. To focus the ions, a positive lens is required, while the aperture lens formed by the accelerator is negative. However, the sign can be reversed using time-dependent accelerating fields. Due to the low temperature of the ions, it is possible to form well-collimated ion beams at very low energies, $U \approx 10-50$ eV. Such ions travel for few microseconds in the accelerator before exiting it. For instance, in the case of $U= 30$ eV, the traveling time is in the order of 3 $\mu$s. This time is long enough to switch the accelerating field while the ions are still in the accelerator.\\
When a constant positive voltage is applied to the anode the ions will experience a negative lens effect. The on-axis longitudinal electric field $E_z(z)$ follows approximately a Gaussian centered on $z_0$, see Fig. \ref{fig:fields} (b).
Fig. \ref{fig:fields} shows the potential applied to the anode $V_a(t)$, the longitudinal electric field component along the beam axis $E_z(z)$ and on-axis radial electric field component $E_r(z)$ for $r \neq 0$, in case of a static (left hand side) or bipolar field (right hand side).
The cylindrical symmetry of the accelerator makes it possible to write the electric field in the accelerator as an expansion of the electric field on the symmetry axis. In a first order approximation, the radial electric field component $E_r(r,z)$ is given by
\begin{equation}
    E_r(r,z) = -\frac{r}{2} \frac{dE_z}{dz},
\end{equation}
where $r$ is the radial position. In the case of a static electric field, the focal strength of the accelerator is described by
\begin{equation}
   \frac{1}{f_0} = - \frac{1}{4} e_z(z_0),
   \label{eq:f_DC}
\end{equation}
where $f_0$ is the time-independent focal length, and the static field  $e_z(z) = E_z(z) / \phi_0$, with the potential given by $\phi_0 = \int_{z_0}^\infty E_z(z') dz'$, as suggested in \cite{Reijnders_JAP_11}. Because the integral of the radial electric field experienced by an ion over time is positive (see Fig. \ref{fig:fields} (c) in dark), a defocussing effect results.\\
By turning off the electric field before the ions leave the accelerator one can cancel the exit kick effect. Moreover, with the use of a bipolar pulse, the exit kick effect can be reversed and a focusing lens created, see Fig. \ref{fig:fields} (f). In this case, the net integral of the radial electric field component will then result to be negative. Typically the electric fields can be changed in less than 100 ns. Bipolar pulses are created with a programmable waveform generator and amplified 50 times. An example of a bipolar pulse $V_a(t)$ used in the experiment can be seen in Fig. \ref{FIG:pulse}.
Here we can define the positive voltage $V_{pos}$ with duration $\tau_p$ and the negative voltage $V_{neg}$ with a duration $\tau_n$ long enough in order that the ions have already left the accelerator when $V_a$ goes to zero. The accelerating pulse is turned on within 100 ns after the ionization laser is off. The ion bunch moves downstream in the direction of the detector (see Fig. \ref{fig:setup}).\\
Now a ``waist scan'' can be performed, i.e. the final spot size is measured depending on one of these parameters. What is actually varied then is the focal strength $1/{f_t}$, which in the case of a bipolar pulse, can be expressed as
\begin{equation}\label{eq:f_bip}
	\frac{1}{f_t}  = \frac{1}{f_0} + \frac{V_{pos} - |V_{neg}|}{4 V_{pos}} e_z(z_s).
\end{equation}
Here, $z_s$ is the position to which the center of the ion bunch has moved when the field is switched from positive to negative (at time $\tau_p$). Obviously, the focal strength depends on $V_{pos}$, $V_{neg}$ and $\tau_p$, not only directly but also through their influence on $z_s$. In the experiment we only vary $V_{neg}$, for a fixed $V_{pos}$ and $\tau_p$. The bunch energy is given by
\begin{equation}\label{eq:en}
U=e \int_{z_0}^\infty{E_z dz}.
\end{equation}
It is of interest to note that $U$ varies during a waist scan due to the different values of $V_{neg}$.\\
For further details about focusing with time dependent fields see Reijnders \textit{et al.} \cite{Reijnders_JAP_11}.

\section{Analysis}
\label{sec:ana}
Images collected from the CCD camera are fitted to a 2-dimensional Gaussian. Due to the fact that the initial volume is not spherical (see Section \ref{sec:setup}), this will result in a non-circular spot on the images. The longer size is by definition in the y-direction and the short size is in the x-direction. Only the x-direction is considered in this paper because along y-direction the bunch suffered from aberrations, which we ascribe to the long beam line and low energy of the ions, which makes them particularly sensitive to small external fields. In the x-direction, residual effects cannot be completely excluded, in which case the extracted temperature will effectively include such distortions, which will generally increase the source emittance.

Particle tracking simulations are performed with the General Particle Tracer code (GPT) \cite{gpt} in order to reproduce the measurements and fit the data. The electric field inside the accelerator has been calculated with the Superfish Poisson solver \cite{Halbach_PA_76}. The initial particle distribution is a 3D Gaussian with dimensions $\sigma_x$, $\sigma_y$ and $\sigma_z$ as listed in Section \ref{sec:setup}, centered at position $z_0$. The initial velocities follow a Boltzmann distribution. A time-dependent $V_a(t)$ is applied depending on parameters varied experimentally, i.e. $V_{neg}$. The \textit{rms} radius of the simulated images are calculated and compared to the experimental data. Minimization of the reduced $\chi^2$ is performed when comparing the measurement data with GPT simulations in order to extract a parameter such as the effective source temperature. According to \cite{Bevington_Book_03}, the accuracy $\sigma_p$ of the parameter that is optimized by minimizing the $\chi^2$ can be calculated from the dependence of $\chi^2$ on that parameter in the region of the minimum as
\begin{equation}
    \sigma_p = \sqrt{2\left(\frac{\partial^2\chi^2}{\partial p^2}\right)^{-1}}.
\end{equation}

\section{Results and discussion}
The measurements presented here consist of waist scans where the negative voltage of the time-dependent pulse $V_{neg}$ has been varied. Because of the change of $V_{neg}$, the focal strength of the time-dependent lens varies, as shown in Eq. (\ref{eq:f_bip}). The other parameters are fixed: $\tau_p = 2$ $\mu$s, $\tau_n =$ 4 $\mu$s (see Fig. \ref{FIG:pulse}) and $V_{pos} =$ 125 V (except in the measurement in Fig. \ref{fig:diffEn}, where this parameter is also varied).\\
The repetition rate during all the experiments is 20 kHz and the exposure time of the CCD is 2 s. Thus every image is the sum of 40000 bunches.

\subsection{Temperature determination} \label{sec:Test}
\begin{figure}[h]
    \centering
        \includegraphics[width=1\linewidth]{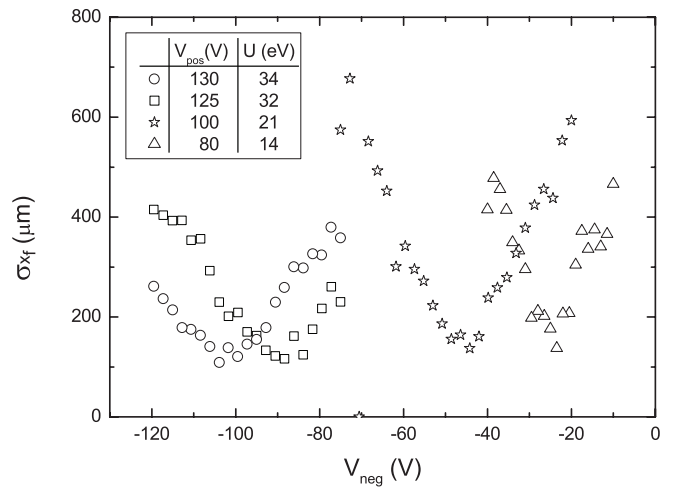}
    \caption{Spot radius $\sigma_{x_f}$ versus negative pulse amplitude $V_{neg}$ for 4 different $V_{pos}$ (indicated in the figure). The bunch energy $U$ at the minimum of the waist scan (which depends on $V_{pos}$ and $V_{neg}$) is also indicated in the label and varies from 14 eV to 34 eV.}
    \label{fig:diffEn}
\end{figure}

\begin{figure}[h]
    \centering
        \includegraphics[width=1\linewidth]{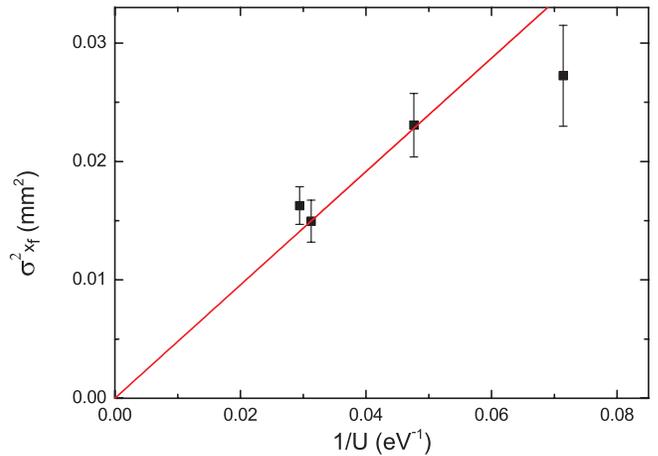}
    \caption{Final spot radius squared $\sigma^2_{x_f}$ versus the reciprocal of the bunch energy $1/U$. The thick line is a linear fit, whose slope is proportional to the effective source temperature $T$.}
    \label{fig:Tdiffen}
\end{figure}

Fig. \ref{fig:diffEn} presents four waist scans at four different $V_{pos}$.
For each $V_{pos}$, the minimum spot radius occurs at different $V_{neg}$, and according to Eq. (\ref{eq:en}) also at a different beam energy $U$. The plot shows that the minimum spot radius $\sigma_{x_f}$ increases when $U$ is lowered: lower energy means longer time of flight and this results in a larger final spot radius. This also indicates that the measurement is indeed sensitive to the effective source temperature. To find $T$, first the minimum spot radius is calculated by fitting the bottom part of a waist with a second order polynomial. The extracted minimum spot radius squared $\sigma_{x_f}^2$ is plotted versus $U^{-1}$ in Fig. \ref{fig:Tdiffen}.
According to Eq. (\ref{eqT}), T can be extracted from a linear fit. The effective source temperature is found to be $T = (4.9 \pm 0.3)$ mK. This temperature, while it is extremely low, is still rather high compared to the expected source temperature ($T_0=143$ $\mu$K), even when $T_0$ is corrected for the fact that the ionization laser beam was turned 0.6 nm above the ionization threshold. In fact, this would make $T_0'=390$ $\mu$K.

\begin{figure}[h]
    \centering
        \includegraphics[width=1\linewidth]{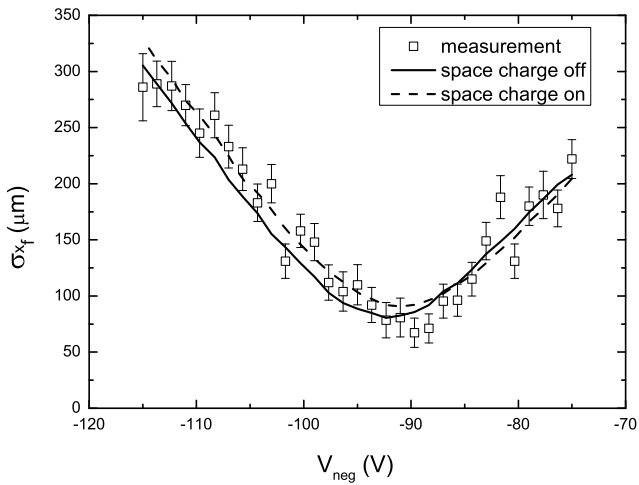}
    \caption{Waist scan performed measuring the final spot size $\sigma_{x_f}$ versus the negative voltage of the time-dependent electric field $V_{neg}$. The experimental data is scattered with the error bars. The solid and the dashed lines are two GPT fits which do and do not include space charge forces respectively.}
    \label{fig:s27_GPT}
\end{figure}

To minimize any effect of space-charge forces on the effective source temperature, we analyzed in detail a waist scan obtained at en even smaller charge of about 0.022 ions per bunch (on average). The data is shown in Fig. \ref{fig:s27_GPT}.
The effective source temperature can be immediately estimated from Eq. (\ref{eqT}), as done in the previous paragraph. Using $U= 32$ eV (the bunch energy at the minimum position) and $\sigma_{x_f} = (73 \pm 4)$ $\mu$m, we find that the resulting effective source temperature $T= (1.8 \pm 0.2)$ mK. This first order approximation is confirmed by a fitting procedure, where the effective source temperature can be extracted more precisely by fitting the behavior of $\sigma_{x_f}$ versus $V_{neg}$ (Fig. \ref{fig:s27_GPT}) with particle tracking simulations. Waist scans are simulated for several source temperatures and the $\chi^2$-minimization procedure from Section \ref{sec:ana} is applied. Fig. \ref{fig:s27_GPT} shows a GPT simulation with parameters that best minimize the $\chi^2$ for different T (solid line). The extracted value is $T = (3 \pm 2)$ mK. The value of $\chi^2 = 1.5$ indicates that the fit is good but could be further improved. %These GPT simulations do not include space charge, which might still be a limiting factor for the source temperature. The measurement shown in Fig. \ref{fig:s27_GPT} are performed with a 4 times smaller charge than the ones shown in Fig. \ref{fig:diffEn} and they result indeed in a lower source temperature. 
%In the next section, we investigate the eventual influence of space charge on the source temperature.\\

\subsection{Influence of space charge}
\begin{figure*}[t]
    \centering
    \subfigure[$n = 1$]{
        \label{n1}
        \includegraphics[width=0.28\linewidth]{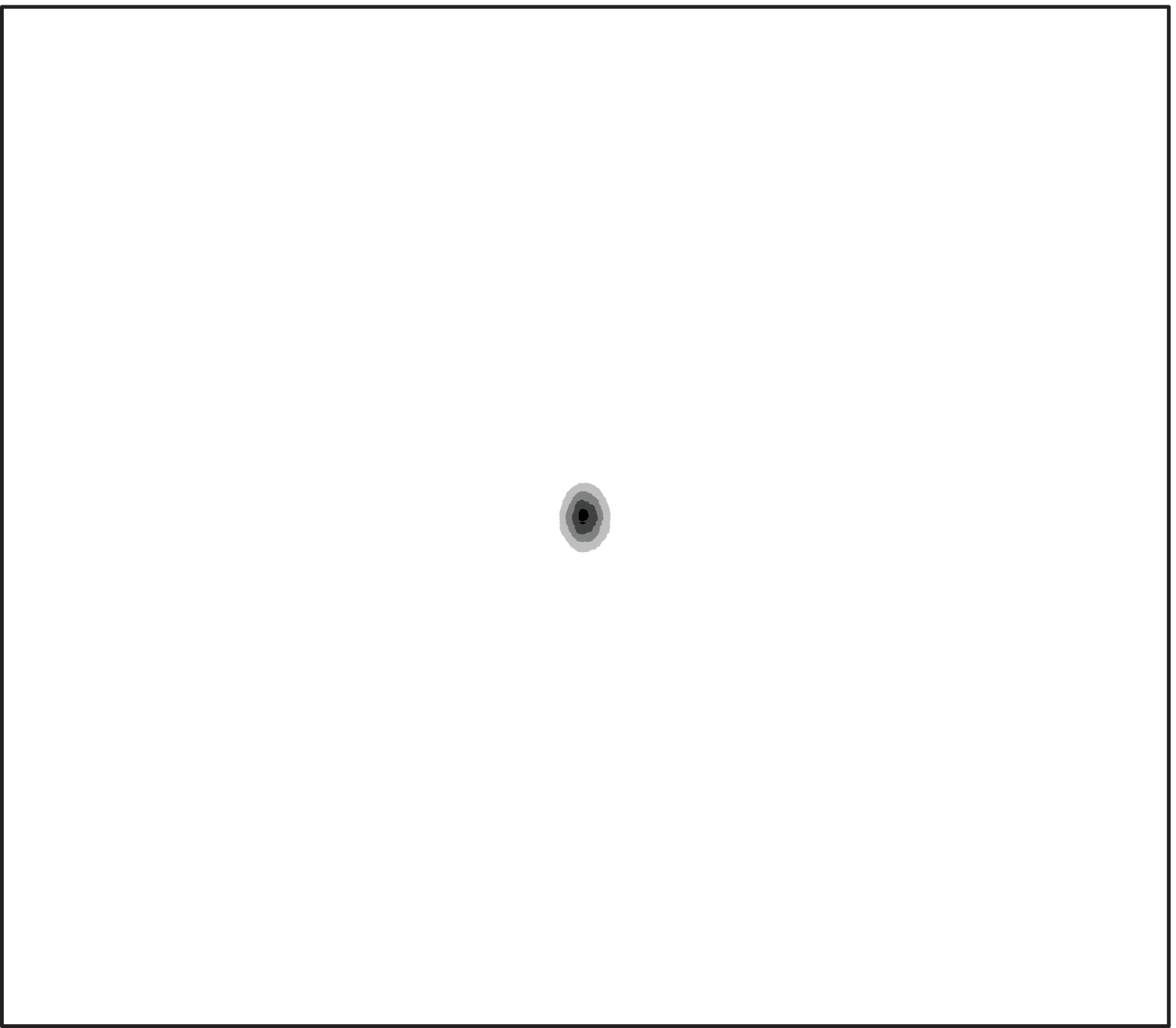}}
    \hspace{.1in}
    \subfigure[$n = 2$]{
        \label{n2}
        \includegraphics[width=0.28\linewidth]{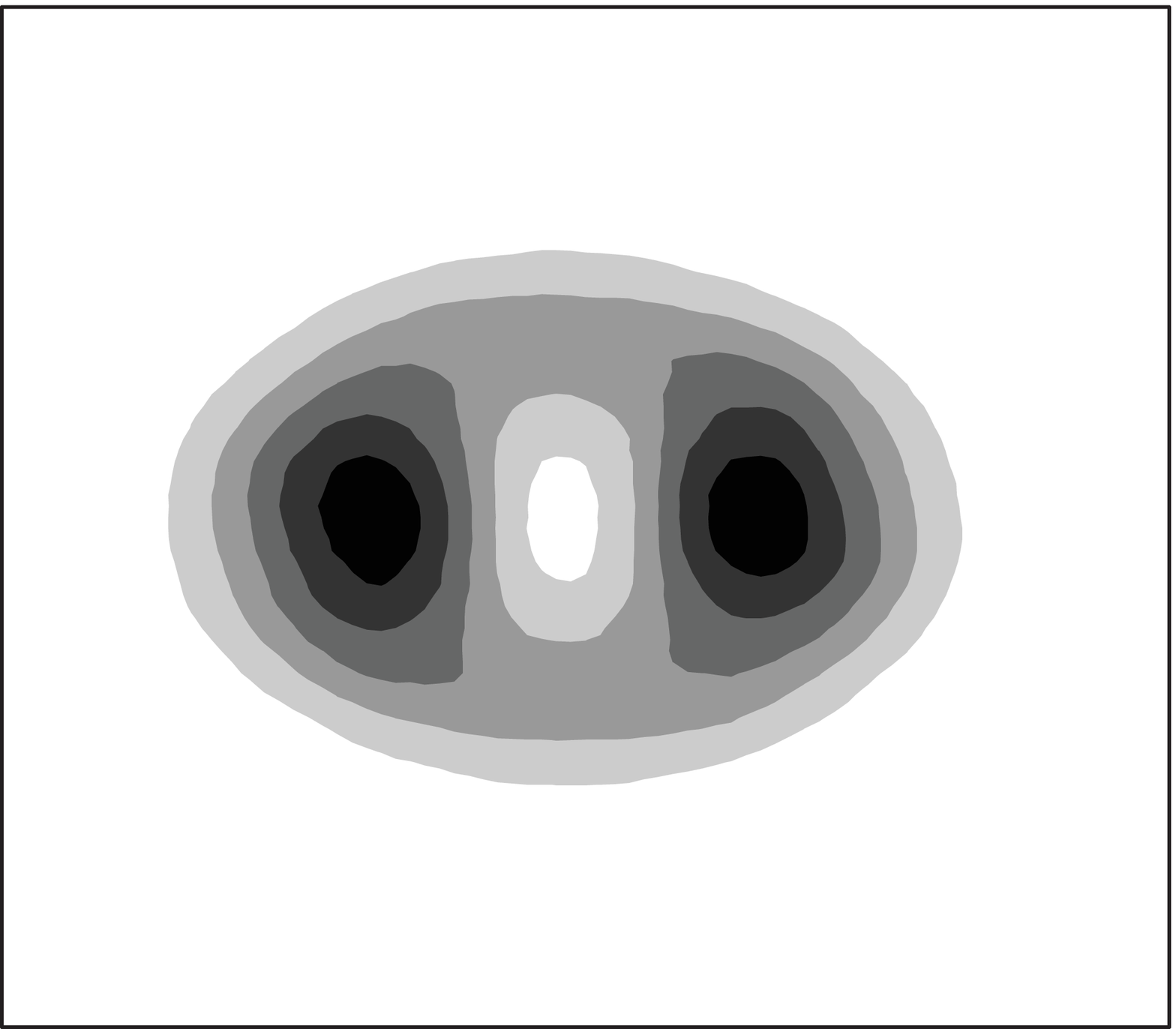}}
    \hspace{.1in}
    \subfigure[$n = 3$]{
        \label{n3}
        \includegraphics[width=0.28\linewidth]{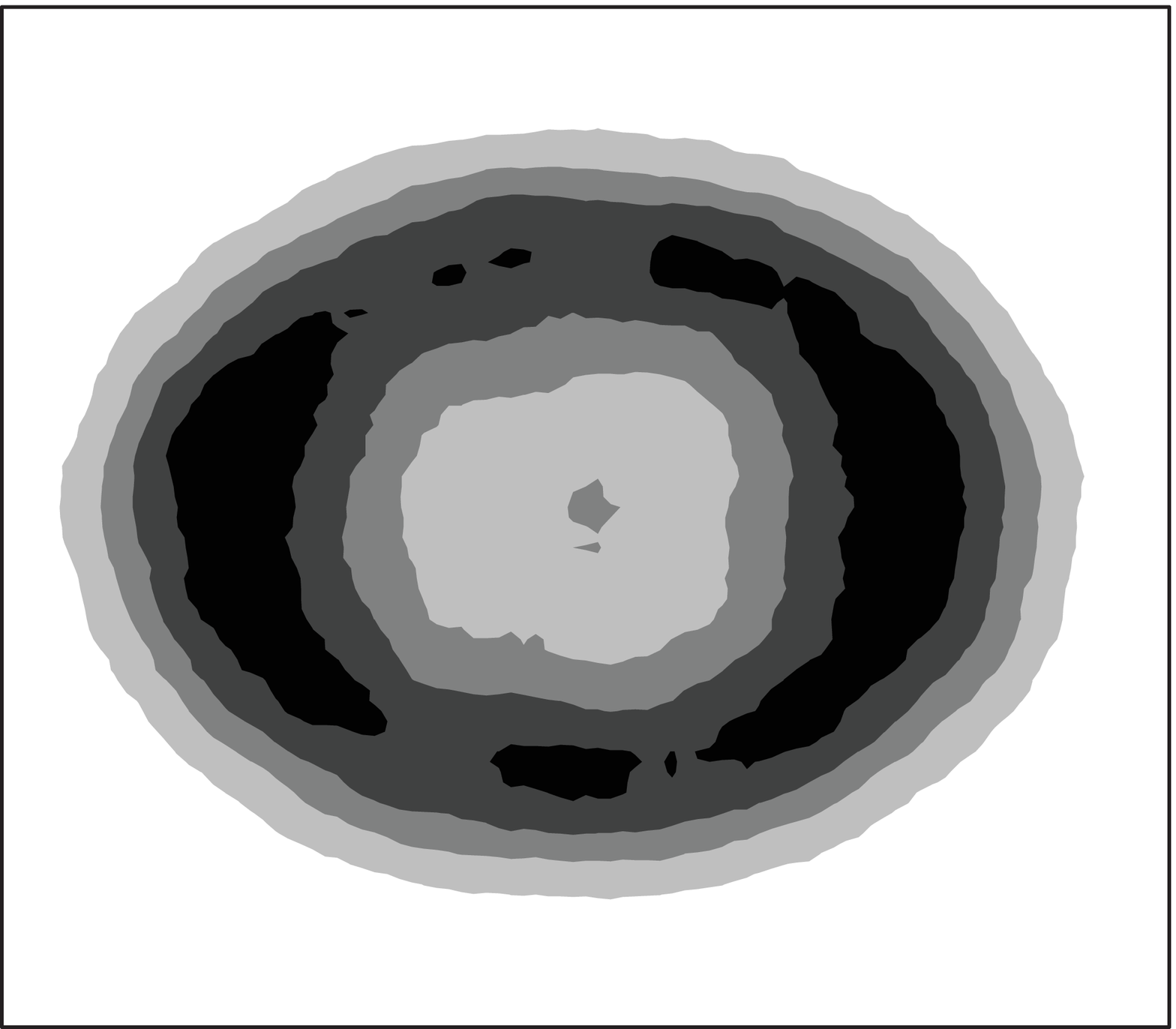}}
    \caption{Particle distribution on the detector in GPT simulations depending on the number of ions in the bunch $n$. The plots are 2D histograms. The scale is the same in the three images: 3.2 by 3.2 mm$^2$. The horizontal axis is the x direction and the vertical axis is the y direction. The \textit{rms} radius for $n = 1,2,3$ in the x-direction is respectively $\sigma_1 = 65$ $\mu$m, $\sigma_2 = 471$ $\mu$m and $\sigma_3 = 635$ $\mu$m. The absolute number of counts per pixel is not of interest here and it is not given but the color code ranges from white (no count) to black (maximum number of counts). The asymmetry present in the spot size is due to the fact that the initial ionization volume is not symmetric.}
    \label{fig:spots}
\end{figure*}

The measurements of Fig. \ref{fig:s27_GPT} were taken using far less than one ion per bunch and high repetition rate. Therefore, space charge forces should not be important in this case; but it is nevertheless interesting to investigate at which level Coulomb interactions can still play a role. The number of ions which are ionized by a laser beam follows Poissonian statistics, so it is possible, even when the expected average number of ions per bunch $\mu$ is less than one, to have some bunches with 2 or more ions. The Poisson distribution is given by $p_n(\mu,n) = (\mu^n e^{-\mu}/n!)$, where $n$ indicates the number of ions per bunch. As an example, when the expected number of ions per bunch is 0.022, the probabilities to obtain one, two or three ions per bunch are respectively 1.96\%, 0.02\% and 0.0001\%. The probability for zero ions is the highest (about 98\%) since only one ionization laser pulse every 50 ionizes one or more atoms. GPT simulations for $n>1$ show a dramatic increase of the spot size due to the low energy of the bunches. This is illustrated in Fig. \ref{fig:spots}, where the simulated distribution at the detector is shown for $n=1$, $n=2$ and $n=3$ at a bunch energy of 32 eV. The \textit{rms} radius for $n = 1,2,3$ in the x-direction is respectively $\sigma_1 = 65$ $\mu$m, $\sigma_2 = 471$ $\mu$m and $\sigma_3 = 635$ $\mu$m. The spot size due to $n=2$ and $n=3$ is much larger than the spot size due to $n=1$ and even if the probabilities for $n=2,3$ are low compared to that for $n=1$ they will play a small role in the final spot size. The asymmetry present in the spots of Fig. \ref{fig:spots} (b) and (c) is due to the fact that the initial ionization volume is not spherical ($\sigma_x < \sigma_y$). In case of three particles per bunch, this effect is still noticeable but more smeared out, see Fig. \ref{fig:spots} (c).

The sum distribution $D(x)$ is given by:
\begin{equation}
    D(x)=\sum_{n=1}^3{n p_n(\mu,n) D_n(x)},
    \label{eq:sum_distr}
\end{equation}
where $D_n(x)$ is the distribution of the final spot for $n= 1,2,3$ obtained from GPT simulations and $p_n$ is the Poisson probability used as a weight. The sum distribution is fitted to a 2-dimensional Gaussian. The standard deviation $\sigma_{sim}$ of the fitted Gaussian along the x-direction at different $V_{neg}$ is also showed in Fig. \ref{fig:s27_GPT} (dashed line). We find that space-charge forces have a very small effect on the waist-scan. The dashed line in Fig. \ref{fig:s27_GPT}, drawn for an effective source temperature of 3 mK, marginally improves the match with the experimental data ($\chi^2 = 1.1$). %seems to represent better the experimental data and it leads to a smaller source temperature of 1 mK, but also leads to a temperature uncertainty on the order of 1 mK.\\
While noticeable, it is uncertain if the resulting broadening can be actually extracted from the data since it is on the order of the accuracy of the measurement. Further investigations are required in order to quantify the influence of space charge.

\section{Conclusion}
We investigated the source temperature of the UCIS, a new kind of ion source that can be used for FIB applications. Time-dependent electric fields are used to focus Rb$^+$ ion bunches and perform waist scans in order to determine the effective source temperature, found to be $(3 \pm 2)$ mK. The 
expected source temperature of $T_0 = 390 \mu$K is consistent with this result given the error bounds; the lower value may point to residual distortions
present in the beam line. The result also weakly depends on space-charge forces. \\
From Eq. (\ref{eq:emitt2}), the \textit{rms} reduced emittance is calculated to be $\epsilon_r = 1.4 \e{-8}$ m rad $\sqrt{eV}$ for an effective source temperature of 3 mK and an initial source size $\sigma_{x_i} = 38$ $\mu$m. The only other emittance measurements of an ultracold ion source was performed by Hanssen \textit{et al.} \cite{Hanssen_NL_08}. They measured the temperature of an ultracold Cr$^+$ source with a different method: observing how the size of an unfocused ion beam increases due to the source temperature when lowering the energy of the beam. The effective source temperature was varied by tuning the ionization laser to a lower wavelength, otherwise the source temperature would be too low to give an appreciable effect on the final spot size. Their measured reduced emittance is a factor 23 smaller because they found a lower source temperature (Chromium has a slightly smaller Doppler temperature\cite{Metcalf_Book_99}) and used an initial source size $\sigma_{x_i} = 5$ $\mu$m. It is possible to create a smaller emittance by reducing the initial size of the source, but the difference in temperature can not be compensated. Due to the uncertainty in our measured source temperature, in fact the two independent measurements overlap within two standard deviations. This confirms that the effective temperature of the UCIS is indeed close to that of the laser-cooled atoms, which is an essential ingredient to achieve high brightness with the UCIS. Moreover, we presented a new method focusing Rb$^+$ bunches with time-dependent fields.\\

\section{ACKNOWLEDGMENTS}
The authors would like to thank J. van de Ven, A. Kemper, W. Kemper, L. van Moll, E. Rietman, I. Koole and H. van Doorn for technical support. This research is supported by the Dutch Technology Foundation STW, applied science division of the "Nederlandse Organisatie voor Wetenschappelijk Onderzoek (NWO)" and the Technology Program of the Ministry of Economic Affairs.

\bibliographystyle{unsrt}
%\bibliography{biblio}

\end{document}